\documentclass[prl,aps,twocolumn,showpacs,amsmath,amssymb,superscriptaddress]{revtex4}
\usepackage{graphicx}
\usepackage{dcolumn}
\usepackage{bm}
\usepackage{subfigure}
\usepackage{setspace}
\usepackage{psfrag}
\usepackage[normalem]{ulem}

\newcommand{\be}{\begin{equation}}
\newcommand{\ee}{\end{equation}}
\newcommand{\ba}{\begin{array}}
\newcommand{\ea}{\end{array}}
\newcommand{\bqa}{\begin{eqnarray}}
\newcommand{\eqa}{\end{eqnarray}}

\newcommand{\ket}[1]{\ensuremath{| #1 \rangle}}

\begin{document}
\title{Microwave driven atoms: from Anderson localization to Einstein's photo effect}
\author{Alexej Schelle}
\affiliation{Physikalisches Institut der Albert-Ludwigs-Universit\"{a}t, 
Hermann-Herder-Str. 3, D-79104 Freiburg, 
Germany}
\affiliation{Laboratoire Kastler-Brossel, 
Universit\'e Pierre et Marie Curie-Paris 6, ENS, CNRS;  
4, Place Jussieu, F-75005 Paris, 
France}
\author{Dominique Delande}
\affiliation{Laboratoire Kastler-Brossel, 
Universit\'e Pierre et Marie Curie-Paris 6, ENS, CNRS;  
4, Place Jussieu, F-75005 Paris, France}
\author{Andreas Buchleitner}
\affiliation{Physikalisches Institut der Albert-Ludwigs-Universit\"{a}t, 
Hermann-Herder-Str. 3, D-79104 Freiburg, 
Germany}
\date{\today}
\begin{abstract}
We study the counterpart of Anderson localization 
in driven one-electron Rydberg atoms. 
By changing the initial Rydberg state 
at fixed microwave frequency and interaction time, 
we numerically monitor the crossover 
from Anderson localization 
to the photo effect in 
the atomic ionization signal.
\end{abstract}
\maketitle
Anderson localization \cite{And, And_exp} is the inhibition 
of quantum transport due to destructive interference
in disordered, static quantum systems. 
When a Hamiltonian quantum system is periodically 
driven and its classical counterpart undergoes a transition to 
chaotic diffusion, 
an analogous localization phenomenon occurs: 
destructive interference between many 
chaotically diffusing trajectories inhibits the 
transport and localizes the diffusing particle's 
wave function \cite{Casbook}. 
Since dynamical chaos rather than static disorder 
establish Anderson's scenario here, the phenomenon is often
labeled {\em dynamical} localization. 

By now, the dynamical variant of Anderson 
localization (and similar phenomena \cite{Wim}) was observed
in a vast range of physical systems -- ranging from 
cold atoms \cite{Raizen94} to photon billiards \cite{Sto} 
and atoms \cite{Rich, Bay, Arn, Gal_Mae}, 
and is best understood in the Floquet or dressed state picture, 
which also allows its formal mapping on 
Anderson's model \cite{Fish}. 
The dressing of the bare system by the driving field photons defines 
multiphoton transition amplitudes between the initial 
and the final field-free state, mediated 
by near-resonantly coupled intermediate states.
These amplitudes need be summed up coherently 
to determine the total transport probability. 
For {\em destructive} interference and thus localization to emerge, 
a {\em large} number of amplitudes is required, what implies 
that the photon energy be {\em small} compared to the energy 
gap between initial and final state. 
This is a scenario in perfect contrast to 
Einstein's photo effect \cite{Ein}, which predicts {\em efficient} 
transport -- mediated by {\em one single} transition amplitude -- for photon 
energies {\em larger} than that energy gap, 
though the general physical context of a driven quantum 
system is identical in both cases. 
Recently, connecting both effects through continuous 
variation of the experimental parameters has moved into 
reach for state of the art atomic physics experiments \cite{recentexpconf},
and it is the purpose of the present Letter to (theoretically) establish this 
connection, and to spell out its characteristic features. 

Our specific atomic physics scenario is defined 
by a one electron Rydberg atom under periodic driving 
by a classical, linearly polarized oscillating electric 
field of amplitude $F$ and frequency $\omega$, described (in length 
gauge and atomic units, employing the dipole approximation) by the 
Hamiltonian 
\begin{equation}
H(t) = \dfrac{\textbf{p}^{2}}{2} - \frac{1}{r} + 
\textbf{F}\cdot\textbf{r}{\rm ~cos}(\omega t)\ ,
\label{Hamiltonian}
\end{equation}
with $\bf p$ and $\bf r$ the electron's momentum and 
position, respectively. 
In this system, quantum transport 
properties are efficiently characterized by the 
ionization probability $P_{\rm ion}(t)$ after a given atom-field interaction 
time $t$, for an atomic initial state $\ket{\Phi_0}=\ket{n_0,\ell_0,m_0}$ 
with well-defined principal and angular momentum 
quantum numbers $n_0$, $\ell_0$ and $m_0$ (the latter one 
being a constant of motion for linearly polarized
driving). 
Transport occurs on the energy axis, from the bound 
initial state towards asymptotically free continuum states, and
is mediated by the absorption of at least 
\begin{equation}
N_{\rightsquigarrow}=\frac{1}{\omega}\left(\frac{1}{2n_0^2} 
- \frac{1}{2n^{2}_{{\rm eff}}}\right)=
\frac{1}{2\omega n^{2}_{0}}\left(1-\frac{n_{0}^{2}}{n^{2}_{{\rm eff}}}\right)
\label{photon_number}
\end{equation}
photons by the electron from the driving field, 
where $n_{\rm eff}<\infty$ denotes the 
{\em effective ionization threshold} 
(at negative energy $-1/2n_{\rm eff}^2$). 
The latter is fixed by the specific experimental conditions and caused by unavoidable
experimental imperfections such as electric stray fields. 
A typical value for state of the art experiments 
is $n_{\rm eff}\simeq 270$ \cite{recentexpconf, Gal_Mae, Gri_Gal_Noe},
which we will employ throughout the sequel of this paper. 
Since, at a given laboratory driving field frequency $\omega=2\pi\times 17.5\ \rm GHz$ 
\cite{recentexpconf}, all field free bound states with $n_{{\rm eff}}>n_{0}\geq230$ will be coupled directly 
to the atomic continuum by {\em one single photon}, such experimental ionization
threshold allows for the continuous interpolation between the 
Anderson limit and the photo effect as described above. 
It suffices to monitor $P_{\rm ion}(t)$ as a function of $n_{0}$, 
with all other experimental parameters fixed. 

We will now model such scan by a faithful 
numerical description of the atomic system under study. 
Our theoretical/numerical tool box is described in detail elsewhere 
\cite{Buc_Del_Gay, Kru_Buc}. 
We only recall here that the theoretical approach combines 
Floquet theory \cite{Shi_per} and complex dilation of the Hamiltonian \cite{Com,Dag}, 
possibly amended by R-matrix theory to account
for the multielectron core of alkali Rydberg states \cite{Kru_Buc}, 
together with considerable computational power provided 
by parallel supercomputing facilities. 
The production of one single data point as displayed 
in the figures below requires repeated diagonalization 
of banded complex symmetric matrices of dimension up to 
$10^{6}$, what amounts to storage needs up to $150$ GB. 

\begin{figure}[t]
\psfrag{xlabel}{$\omega n_{0}^{3}$}
\psfrag{ylabel}{$F^{10{\%}}n_{0}^{4}$}
\psfrag{ARR}{(I)}
\psfrag{FP}{(II)}
\psfrag{ER}{(III)}
\psfrag{photo}{photo effect}
\psfrag{loc}{\hspace{-.04cm}localization}
\psfrag{N1}{\begin{small}$N_{\rightsquigarrow}=1$\end{small}}
\psfrag{N2}{\begin{small}$N_{\rightsquigarrow}=2$\end{small}}
\psfrag{N3}{\begin{small}$N_{\rightsquigarrow}=3$\end{small}}
\includegraphics[width=8.6cm, height=5cm, angle=0.0]{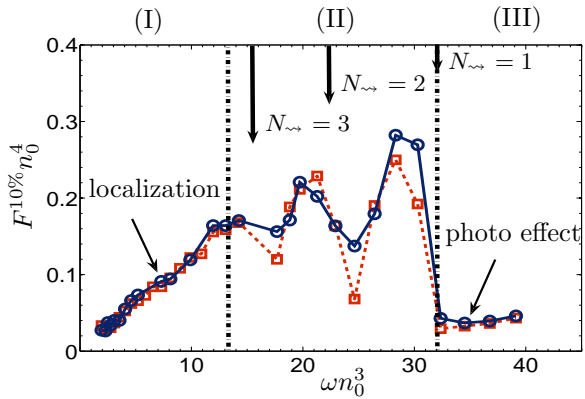}
\caption{(color online) Scaled ionization threshold field 
$F_{0}^{10{\%}}=F^{10{\%}}n_{0}^{4}$ of atomic 
hydrogen (red dashed line ($\square$)) and lithium (blue solid line (O)),
at fixed laboratory microwave frequency 
$\omega=17.5$ GHz and interaction time $t=500$ ns. 
The scaled frequency $\omega_{0}=\omega n_{0}^3 = 1.9...39.1$ is tuned by 
changing the initial state's principal quantum number from $n_{0}=90$ to 
$n_{0}=245$, at fixed values of the angular 
momentum quantum numbers $\ell_{0}=1$ and $m_{0}=0$. 
We observe three distinct regimes:
(I), $1.9\leq\omega_0\leq 13.1$: the monotonous increase of $F_{0}^{10{\%}}$ 
with $\omega_{0}$ is a characteristic signature
of Anderson localization in strongly driven quantum systems \cite{Kru_Buc2}.
Regime (II),  $13.1\leq\omega_{0}<31.5$: $F_{0}^{10{\%}}$ still 
increases with $\omega_0$, on average, but is garnished by large 
modulations due to the passage of the 
atomic initial state across subsequent 
$N_{\rightsquigarrow}$-photon ionization thresholds indicated by vertical arrows. 
Anderson localization and finite $N_{\rightsquigarrow}$-photon ionization coexist.
Regime (III),  $\omega_{0}\geq31.5$: The photon energy exceeds the 
ionization potential of the initial state, the Anderson scenario 
is inapplicable, and single photon absorption mediates the ionization process.}
\label{scaled_field}
\end{figure}

In order to highlight the continuous transition 
from suppressed transport due to 
Anderson localization to enhanced transport due 
to the photo effect, we scan an energy range
of atomic initial states from $n_0=90$ to $n_0=245$, 
at fixed microwave frequency and atom-field 
interaction time $t=500\ \rm ns$, and angular momentum 
quantum numbers $\ell_0=1$, $m_0=0$. 
The specific choice of these parameters is inspired by 
ongoing experiments \cite{recentexpconf} on Rydberg states of 
lithium, and we will provide data for lithium as well 
as for atomic hydrogen, to disentangle universal 
features of the said transition from those characteristic 
of the atomic species under scrutiny.
Furthermore, our ``starting value" $n_0=90$ guarantees that 
we start out in the Anderson regime, where the
ionization yield is characterized by a universal 
ionization threshold {\em irrespective} of the 
atomic species \cite{Kru_Buc2}.

Figure~\ref{scaled_field} shows the results of our 
calculation in terms of the 
{\em scaled ionization threshold field} $F^{10{\%}}_{0}=F^{10\%}n_0^4$,
i.e., of the driving field amplitude $F_0^{10\%}$ which 
induces $P_{\rm ion}(t)=0.1$, measured in units of the 
Coulomb field experienced by the electron on its 
unperturbed Rydberg orbit $n_0$ \footnote{Our numerical threshold fields 
are extracted from the $F$-dependence of the ionization yield as the fundamental experimental observable \cite{Buc_Del_Gay,Kru_Buc}. For all 
other parameters 
fixed, the relative numerical error  $\Delta F^{10{\%}}_{0}$ is determined by the step size in $F$, 
and is estimated on the order of few percent. This -- as well as the weak dependence 
of  $F^{10{\%}}_{0}$ on $n_{\rm eff}$ (which was fixed  at $n_{\rm eff}=270$ in our present calculations) -- leaves the significance of  the structures observed 
in Figs.~1-3 qualitatively unaffected  (for a related discussion, see also \cite{Buc_Del95}).}. 
The threshold field is plotted as a function of 
the {\em scaled driving field frequency} 
$\omega_0=\omega n_0^3$, i.e., of the driving field 
frequency $\omega$ measured in units of the unperturbed 
Kepler frequency for $n_0$. 
Clearly, we can identify three regimes of 
qualitatively different behavior: In regime (I), for 
low principal quantum numbers $n_0=90\ldots 170$  
(corresponding to scaled frequencies $\omega_0\simeq1.9\ldots 13.1$), 
we witness the characteristic signature
of Anderson localization -- the scaled ionization threshold {\em increases} 
with the excitation of the initial atomic state, i.e., with {\em decreasing} 
ionization potential, and is essentially independent of the 
atomic species \cite{Kru_Buc2}.
In regime (II), the ionization threshold 
still increases on average -- suggestive 
of Anderson localization -- but is garnished by large-scale modulations.
Closer inspection of this oscillating structure reveals its origin in successive
passages through the multiphoton ionization thresholds indicated by vertical arrows 
in the figure: 
The opening of 
a direct, $N_{\rightsquigarrow}$-photon ionization channel \cite{Faisal}
is manifest in a local, rapid decrease of $F_0^{10\%}$ with $\omega_0$ 
(since the dominant 
contribution to the ionization signal is of lower order). As $\omega_0$ increases further on, the
threshold field increases again, since the cross section for $N_{\rightsquigarrow}$-photon ionization 
decreases with increasing frequency -- until the next channel opens.
The thus emerging structures are precursors of the final opening of the 
single photon ionization channel at $n_0=230$ ($\omega_0=32.4$),
which defines the demarkation line between regime (II) and the 
realm of the photo effect, (III) \footnote{We verified that our numerical thresholds in regime (III) deviate
from semiclassical estimates of the one-photon threshold \cite{Cas} by less than $30\%$. The discrepancy is due to the finite 
value of $n_{\rm eff}<\infty$.}.

We therefore witness a synchronicity of Anderson 
localization and ($N_{\rightsquigarrow}$-order) photo effect 
in regime (II): the former still largely suppresses 
the ionization process, even when, 
by virtue of the value of $N_{\rightsquigarrow}$,
multiphoton transitions of very low
order mediate the transport, while the latter is already reflected 
in prominent non-monotonicities of the threshold field. Only in regime (III) is Anderson localization 
completely absent. 
\begin{figure}[t]
\psfrag{AR}{(I)}
\psfrag{FP}{(II)}
\psfrag{ER}{(III)}
\psfrag{xlabel}{$\omega n_{0}^{3}$}
\psfrag{ylabel}{$\xi(\omega_{0},F^{10{\%}}_{0})/N_{\rightsquigarrow}$}
\includegraphics[width=8.6cm,height=5cm,angle=0.0]{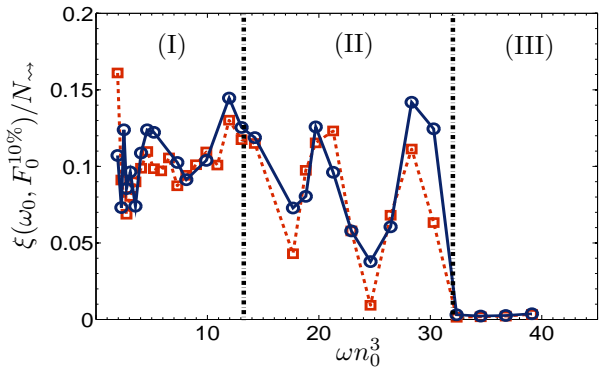}
\caption{(color online) Ratio of the atomic localization 
length $\xi$ as estimated by Eq.~(\ref{localization_length}), 
to the number of absorbed 
photons $N_{\rightsquigarrow}$,
vs. scaled frequency $\omega n_{0}^{3}$. Data are extracted from the $10\%$- 
ionization thresholds of Fig.~1, for atomic hydrogen (red dashed line ($\square$)),
and 
lithium (blue solid line (O)).
On average, $\xi/N_{\rightsquigarrow}$
is 
constant in regime (I)
-- a hallmark of exponential 
localization of the electronic wave function on the energy axis.  This is still 
true in regime (II), where, however, the discreteness of the lattice (on the energy axis) strongly 
affects the transport behavior: 
Large scale modulations of the signal emerge due to direct $N_{\rightsquigarrow}$
photon transitions to the continuum. 
Only in regime (III) does $\xi$ drop to zero, 
thus invalidating the Anderson picture.}
\label{localization_length}
\end{figure}

A complementary analysis corroborates 
this interpretation. According to the theory of Anderson
localization, the exponential localization of the electronic 
wave function on a characteristic scale $\xi$ (in units of the driving field 
photon energy $\hbar\omega$) on the energy axis \cite{Cas} implies an exponential 
scaling of the ionization yield, according to 
$P_{\rm ion}\sim\exp(-2N_{\rightsquigarrow}/\xi)$.
Consequently, for a fixed ionization yield (as implicit in 
the definition of $F_0^{10\%}$), this leads to the prediction that  
$\xi(\omega_{0},F^{10{\%}}_{0})/N_ {\rightsquigarrow}$ 
be {\em independent} of $\omega_{0}$. 
This is what is observed in 
Fig.~\ref{localization_length} in regime (I) (modulo threshold fluctuations which are 
characteristic for the Anderson problem \cite{abu98}),
where we plot 
$\xi(\omega_{0},F^{10{\%}}_{0})/N_ {\rightsquigarrow}$ 
vs. $\omega_0$, with $\xi$ estimated according to \cite{Cas}: 
\begin{equation}
\label{eq:xi}
\xi \simeq 3.33F_{0}^2\omega_{0}^{-10/3}n_{0}^{2}\ .
\end{equation}
This simple expression is known to be quantitatively 
incorrect \cite{Buc_Del_Gay,Gal_Mae}, but to provide a qualitatively reliable 
characterization of the general trend of $\xi$
with $n_{0}$. Furthermore,
Fig.~2 clearly spells out that exponential 
localization of the electronic 
wave function on the energy axis prevails, at least on average, even deeply into 
regime (II), where the ionization behavior is simultaneously 
strongly affected by the described opening
of few-photon ionization channels. 
In the Anderson picture, the latter is tantamount of 
finite size effects which manifest in localization lengths 
$\xi$ of order unity, thus resolving the granularity of the lattice 
(on the energy axis) along which transport occurs. 
In regime (III), the lattice constant (i.e., here, the photon energy \cite{Fish, Cas})
is larger than the effective sample length, 
and the Anderson picture turns inapplicable. 

Let us finally analyze the characteristics of the atomic 
transport process in terms of its ``complexity", which 
can be characterized in terms of the number of Floquet 
eigenstates which mediate the ionization process -- what in turn 
provides a measure of the volume of Hilbert space which 
is effectively explored in the course of the ionization process. 
A good estimate thereof is given by the Shannon width \cite{Shannon}
\begin{equation}
\mathcal{W}(F^{10{\%}}_{0},\omega_{0}) = 
{\rm exp}\left[ -\sum_{j} |w_{j}|^2\ln |w_{j}|^2\right]\ ,
\label{Shannon}
\end{equation}
which raises the Shannon entropy of the 
decomposition $\ket{\Phi_0}=\sum_jw_j\ket{\epsilon_j}$ 
of the atomic initial state in the 
Floquet basis $\lbrace\ket{ \epsilon_j}\rbrace$, with individual 
weights $|w_j|^2$ \cite{abu98}, to the number of Floquet states which effectively contribute. 
According to our qualitative understanding of Anderson 
localization on the one hand and the photo effect on the 
other, the former is the consequence of the destructive 
interference of a {\em large} number of multiphoton transition amplitudes,
while the latter is mediated by essentially {\em one single} transition matrix element. 
Consequently, Anderson localization implies
the coupling of a large number of states, and this is 
tantamount of the spreading of the field-free atomic initial state over
a large number of atomic eigenstates {\em in} the field, 
whereas in the photo effect the atomic initial state is directly coupled 
to the continuum, without the participation of other bound states. 
Correspondingly, the Shannon width should be large in one
case, and small, rather close to one, in the other.
\begin{figure}[b]
\psfrag{AR}{(I)}
\psfrag{FP}{(II)}
\psfrag{ER}{(III)}
\psfrag{xlabel}{$\omega n_{0}^{3}$}
\psfrag{ylabel}{$\mathcal{W}(F^{10{\%}}_{0},\omega_{0})$}
\includegraphics[width=8.6cm,height=5cm,angle=0.0]{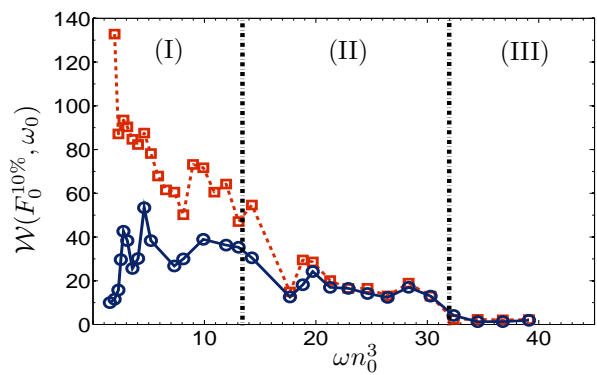}
\caption{(color online) Shannon width, Eq. (\ref{Shannon}),
at the $10\%$-ionization threshold, as a function 
of the scaled frequency $\omega_0=\omega n_{0}^{3}$, for the
 same parameters as in Fig.~\ref{scaled_field}.
 Remarkably, $\mathcal{W}$ takes appreciable values for initial
 states up to right below the single photon ionization threshold at the lower
 edge of regime (III). 
 In regime (I), atomic hydrogen exhibits considerably larger 
 values than lithium, what we attribute to the angular momentum degeneracy 
 of the hydrogenic initial state.}
\label{Shannon_fig}
\end{figure}
Figure~\ref{Shannon_fig} confirms this expectation: As a function of 
$\omega_0$, at fixed laboratory value $\omega$, the Shannon width 
exhibits large values in regime (I), and decreases almost 
monotonically to a level close to unity in regime (III), with 
intermediate values around approx. 20 in regime (II). 
Much as in our previous analysis 
of the ionization threshold's and the localization 
length's dependence on the scaled frequency it is 
also here evident that the interference of 
multiple transition amplitudes as the fundamental 
mechanism of Anderson localization prevails very far 
into regime (II), even in the presence of 
already relatively efficient direct continuum coupling through 
few-photon ionization channels. 
The relatively high level of $\mathcal{W}\simeq 20$ even at $\omega_0\simeq 30$
convincingly demonstrates the rapid proliferation of multiphoton 
coupling amplitudes as the photon energy becomes smaller than the initial state's 
ionization potential. 
In terms of the Anderson model, even small lattices suffice for 
the emergence of Anderson-like suppression of diffusive transport. 

Figure~\ref{Shannon_fig} also highlights some subtle 
differences of the ionization process for different 
atomic species -- here lithium and atomic hydrogen:
In regime (I), $\mathcal{W}$ is significantly larger 
for hydrogen than for lithium, what we attribute to 
the higher degeneracy of the hydrogen atom's 
initial state's angular 
momentum manifold as compared to the non-hydrogenic 
initial state of lithium ($m_0=0$ renders $\ell=0$ 
directly accessible by single photon absorption from 
$\ell_0=1$) \cite{Blue}. 
The progressive vanishing of this discrepancy 
in regime (II) is consistent with the reduction of $N_ {\rightsquigarrow}$.

In summary, we established a continuous transition 
from Anderson localized quantum transport to 
the photo effect, by simple tuning of the sample 
length at fixed lattice constant, which, in our specific,  
experimentally relevant example from atomic physics, are 
defined by the ionization potential of the atomic initial 
state $\ket{n_0,l_0,m_0}$ and 
the energy $\hbar\omega$ of the injected photons, respectively. 
We have seen that both transport mechanisms coexist in a 
certain parameter range, where Anderson localization is 
garnished by finite size effects, which, in an atomic physics language,
are nothing but the opening of multiphoton ionization channels. 
Our most remarkable observation is probably that 
characteristic signatures of Anderson localization 
prevail in the ionization signal even when absorption 
of very few photons suffices to ionize
the Rydberg electron: Thus, quasi-randomness as a necessary 
prerequisite of Anderson localization is rapidly established,
if only the local spectral density permits the coupling of 
many unperturbed atomic states by comparably few photons.

During the preparation of this manuscript, we have learned
that the predicted ionization behavior in regime (II) was recently 
observed experimentally \cite{recentexpconf}. 

We thank Joshua Gurian, Haruka Maeda and Tom Gallagher 
for helpful discussions, and
LRZ at 
the Bavarian Academy of Sciences,
RZG of the Max Planck Society, 
and IDRIS Paris for providing computation time. A.S. and A.B. are grateful for 
partial funding through DFG (Forschergruppe 760), and 
A.S. acknowledges financial support 
through the QUFAR Marie Curie Action MEST-CT-2004-503847.

\end{document}